\documentclass[prlb,aps,amssym,nofootinbib,floatfix,reprint,notitlepage]{revtex4-2} 

\usepackage{xcolor}
\usepackage{amsmath,amssymb,bbold,bm}
\usepackage{graphicx}
\usepackage[normalem]{ulem}
\usepackage{cancel}
\usepackage{comment}

\usepackage[pdfpagelabels,colorlinks=true,breaklinks=true,linktocpage=true,linkcolor=red,urlcolor=blue,citecolor=blue,%
]{hyperref}

\newcommand{\sbraket}[1]{\langle #1 \rangle}
\newcommand{\be}{\begin{equation}}
\newcommand{\ben}{\begin{equation*}}
\newcommand{\ee}{\end{equation}}
\newcommand{\een}{\end{equation*}}
\newcommand{\bs}{\begin{split}}
\newcommand{\es}{\end{split}}
\newcommand{\bmx}{\begin{array}}
\newcommand{\emx}{\end{array}}

\newcommand{\bea}{\begin{eqnarray}}
\newcommand{\bean}{\begin{eqnarray*}}
\newcommand{\eea}{\end{eqnarray}}
\newcommand{\eean}{\end{eqnarray*}}
\newcommand{\dg}{^{\dagger}}
\newcommand{\dn}{^{\vphantom{\dagger}}}

\newcommand{\ua}{\uparrow}
\newcommand{\da}{\downarrow}

\newcommand{\bb}[1]{\mathbb{#1}}

\newcommand{\eps}{\epsilon}

\newcommand{\pref}[1]{(\ref{#1})}

\newcommand{\intob}[1]{\int_{0}^{\beta}{#1}}

\newcommand{\abs}[1]{\left\vert #1 \right\vert}

\newcommand{\ket}[1]{\left\vert #1\right\rangle}
\newcommand{\braket}[1]{\left\langle #1\right\rangle}

\newcommand{\mat}[1]{\left(\bmx{cc}#1\emx\right)}

\newcommand{\bw}[1]{\begin{widetext}}
\newcommand{\ew}[1]{\end{widetext}}

\newcommand{\gray}[1]{}

\newcommand{\nothing}[1]{}

\begin{document}

\title{Dynamic-RKKY induced time-reversal symmetry breaking and chiral spin liquids}
\author{Siqi Shao\,$^{1}$}
\author{Yang Ge$^{1,2}$}
\author{Yashar Komijani\,$^{1*}$}
\affiliation{$^1$Department of Physics, University of Cincinnati, Cincinnati, Ohio 45221, USA}
\affiliation{$^2${Department of Physics and Engineering, Tulane University, New Orleans,  Louisiana 70118, USA}}
\date{\today}
\begin{abstract}
We study the Ruderman–Kittel–Kasuya–Yosida (RKKY) interaction in various Kondo lattice systems. We argue that the weak Kondo coupling expansion contains certain physics which is lost in the usual static approximation to the spin susceptibility. Most notably, while the former is sensitive to the time-reversal symmetry breaking, the latter is blind to it. Using exact diagonalization on small systems, we show that this enables inducing spin chirality by an external magnetic field. To study larger systems, we use a large-N approximation to capture the effect of dynamic-RKKY interaction on U(1) spin liquids. On a honeycomb Kondo lattice with Haldane fluxes for electrons, we show that the topology and chiral edge states are induced on the spinons. Our results suggest that dynamic RKKY in combination with external magnetic field or in proximity to topological electronic materials, can be used as a tunable Dzyaloshinskii-Moriya even in centrosymmetric materials.
\end{abstract}
\maketitle
\section{Introduction}
In recent years, the study of magnetic interactions in Kondo lattice systems has gained significant attention due to its relevance to quantum spin liquids \cite{Fradkin2013,Savary2016,Zhou2017,Knolle2019,Broholm2020}, heavy fermions \cite{Coleman2015}, and topological materials \cite{Hasan2010,Qi2011,Armitage2018}. Among these, the Ruderman–Kittel–Kasuya–Yosida (RKKY) interaction  \cite{Kasuya1956} plays a central role in mediating indirect exchange coupling between localized spins through conduction electrons. Recent studies have explored various aspects of the RKKY interaction in complex systems, including in relation to intrinsic frustration in non-Fermi-liquid metals \cite{she2013} as well as in the emergence of quantum spin liquid phases in RKKY-coupled systems \cite{wojcik2023}. In these non-centrosymmetric candidate materials for quantum spin-liquid, often a significant Dzyaloshinskii-Moriya interaction is present which induces \cite{kang2024, he2024} chirality and topology to the underlying spinons. Other research investigates the interplay between RKKY and Kondo interactions in helical Luttinger liquids, offering new insights into their combined effects \cite{lee2015,yevtushenko2018}. Additionally, higher-order effects, such as ring exchange processes, have been analyzed, further expanding our understanding of RKKY interactions in these systems \cite{zitko2024}.

The conventional treatment of RKKY interaction often involves a static approximation of the spin susceptibility \cite{kogan2011}, which overlooks dynamic effects that may become crucial in various physical regimes.  In this work, we revisit the RKKY interaction in the weak Kondo coupling regime, extending the conventional analysis to include dynamic effects. We demonstrate that the dynamic RKKY interaction, derived from time-dependent fluctuations of the conduction electron spin susceptibility, retains certain key physical phenomena—such as sensitivity to time-reversal symmetry (TRS) breaking—that are lost in the static approximation. Specifically, we show how this dynamic coupling enables the induction of spin chirality by external magnetic fields or through proximity to topologically nontrivial materials.

The dynamic-RKKY coupled spins is a strongly correlated problem. Using exact diagonalization on small systems and a large-N approximation \cite{Arovas1988,Read91} on extended lattices, we explore the impact of dynamic RKKY interaction on spin liquids and spinons. Notably, our modified large-N treatment includes a novelty that makes the formalism (not the result) agnostic to whether the interaction channel is attractive or repulsive. This element is crucial for capturing the dynamic RKKY interaction whose strength and sign changes in space and time. Using this method, on a honeycomb Kondo lattice with Haldane fluxes, we show that dynamic RKKY can induce topological chiral spin states, generating non-trivial edge modes. This tunable interaction provides a mechanism to engineer Dzyaloshinskii-Moriya-like effects, even in centrosymmetric systems, opening avenues for the design of chiral spin liquids and topologically ordered states in Kondo lattices.

\begin{figure}[tp]
\includegraphics[width=\linewidth]{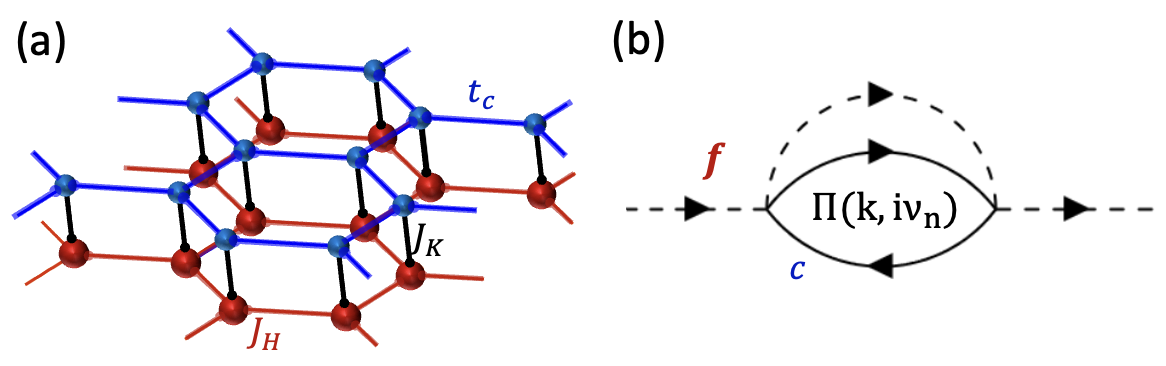}
\caption{(a) Typical Kondo lattice setups studied in this paper. (b) The dynamic-RKKY contribution to the spinon self-energy is due to twice scattering of the spinons $f$ off the electron gas and involves spin-susceptibility or vacuum spin polarization $\chi(k,i\nu_n)$ of conduction electrons $c$.}\label{fig1}
\end{figure}

The rest of the paper is as follows. The model and the method are discussed in section \pref{sec:method} including symmetry analysis and large-N approximation. The results on Kondo triangle and Kondo lattice are presented in section \pref{sec:results}. This also includes a discussion of the RKKY-induced spinon delocalization temperature. The paper is concluded in section \pref{sec:conclusion}, offering a list of open questions and future directions. Finally, a number of appendices provide proof of statements or generalization of the results of the paper.
\section{Method}\label{sec:method}
In the paper, we study the Kondo-Heisenberg model in the weak Kondo coupling limit. The Hamiltonian has three parts:
\bea
H&=&\sum_{ k} \eps_{ k} c^\dagger_{ k} c\dn_{ k}
+\sum_{{ij}}J_H^{{ij}} \vec S_{i}\cdot\vec S_{ j}
+\sum_{ j}  J_K \vec S_{ j}\cdot c_{ j}^{\dagger}\vec \sigma c\dn_{ j}
\eea
Here, $c=\mat{c_\ua & c_\da}^T$ are spinors, $ k\in {\rm BZ}$ is the crystal momentum and $ i,j$ are real-space indices of sites in a lattice in arbitrary dimensions, related by $c_{ i}=\frac{1}{\cal N}\sum_{ k}e^{i{ k}\cdot{ j}}c_{ k}$ where ${\cal N}$ is the number of sites in the system.

For simplicity and without loss of generality, we have assumed that the Kondo interaction is local which is not applicable to topological Kondo insulators \cite{dzero2010topological,Dzero2016}. For future references, we refer the last term of the Hamiltonian with the $J_K$ coupling as the Kondo interaction $H_K$. In the following, we discuss the weak Kondo coupling limit, relevant to this paper.

\subsection{Dynamic RKKY Interaction}
In the weak coupling regime, $J_K$ remains small compared to other energy scales and can be treated perturbatively. In this case, we can integrate out conduction electrons by expanding the action to second order in $J_K$.
\be
e^{-S_{\rm eff}[\vec S]}=\int{D(\bar c,c)e^{-S[\bar c,c,\vec S]}}
\ee
We find (see Appendix\ref{sec:app0}) the effective action $S_{\rm eff}=S_B+S_{\rm int}$, where $S_B$ is the Berry term of the spins and $S_{int}$ given by
\be
S_{\rm int}=\int{d\tau_1d\tau_2}J^{\rm eff}_{{ij}}(\tau_1,\tau_2)\vec S_{ i}(\tau_1)\cdot\vec S_{ j}(\tau_2)\label{eqS}
\ee
where the effective interaction between spins is
\be
J^{\rm eff}_{ {ij}}(\tau_1,\tau_2)=J^H_{{ij}}\delta(\tau_1-\tau_2)+\frac{1}{2}J_K^2\chi_{ij}(\tau_1-\tau_2)\label{eqJeff}
\ee
The first term is the usual instantaneous Heisenberg interaction while the later term is the dynamic RKKY interaction. Here,
\be
\chi_{ij}(\tau_1,\tau_2)=\sbraket{-T_\tau (c\dg \vec\sigma c\dn)_{i,\tau_1}\cdot(c\dg\vec\sigma c\dn)_{j,\tau_2}}_0
\ee
is the bare spin susceptibility or spin-sector vacuum polarizability of the conduction electrons, evaluated in absence of Kondo interaction. 

The action \,\pref{eqS} and \pref{eqJeff} is the dynamic RKKY problem, introduced and treated here. The usual static approximation is obtained by assuming the spins vary slowly in time $\vec S_i(\tau_1)\sim \vec S_i(\bar\tau)$ and $\vec S_j(\tau_2)\to \vec S_j(\bar\tau)$, as a result of which the action \pref{eqS} reduces to
\be
S_{\rm int}\approx \intob{d\bar\tau}\vec S_i(\bar\tau)\vec S_j(\bar\tau) J_{ij}^{\rm static}
\ee
in terms of an effective $J_{ij}^{\rm static}$ which involves static (zero frequency) spin-susceptibility $\tilde\chi_{ij}(0)$:
\be
J_{ij}^{\rm static}\equiv\int{d\Delta\tau} J_{ij}^{\rm eff}(\Delta\tau)=J^H_{{ij}}+\frac{1}{2}J_K^2\tilde\chi_{ij}(0).\label{static1}
\ee
Such an approximation might seem harmless in gapped systems when retardation effects are negligible but as discussed below, this is not true. Another simple limit is the instantaneous approximation, in which $J_{ij}^{\rm instan}(\tau_1,\tau_2)\equiv [J_{ij}^{H}+J_K^2\chi_{ij}(0)]\delta(\tau_1,\tau_2)$. As we explain below certain physics is lost in both of these approximations but retained in the dynamic RKKY interaction. We aim to capture these physics and to this end, the original equations \pref{eqS} and \pref{eqJeff} have to be solved.

For simplicity, we assume conduction electrons to be in a paramagnetic state and without any spin-orbit coupling. This means that $\chi$ is the same as the charge susceptibility (charge-sector vacuum polarizability) $\chi_{ij}(\tau)=\Pi_{ij}(\tau)$.  If additionally, the conduction electrons are non-interacting, we can apply Wick's theorem to reduce it to the Lindhard function \cite{wen2004,Coleman2015} $\Pi_{ij}(\tau)=g_{ij}(\tau)g_{ji}(-\tau)$, 
expressed in terms of the conduction electron propagators $g_{ij}(\tau)=\sbraket{-T_\tau c\dn_i(\tau)c\dg_j}$. 
Next we study symmetry properties of these equations.

\subsection{Symmetry analysis}
In this section we analyze the consequence of various symmetries for the electronic Green's function and the Lindhard function, which governs the properties of $J^{\rm eff}(\tau)$. Here we only list the symmetries and defer the proofs to the Appendix \ref{sec:app1}.

\emph{Translational symmetry} -- In bravais lattices with translational symmetry $g_{ij}$ and $\Pi_{ij}$ are only functions of $i-j$. 
In general non-bravais (e.g. bipartite) lattices, the Lindhard function has the structure
\be
\Pi_{aa'}(n,\tau)=g_{aa'}(n,\tau)g_{a'a}(-n,-\tau)\label{eqP}
\ee
in terms of $g_{aa'}(n,\tau)=\sbraket{-T_\tau c\dn_a(n,\tau)c\dg_{a'}(0)}$. For example in a honeycomb lattice considered here we have $a,a'=A,B$. We define the Fourier transform $\Pi_{aa'}(q,z)$ with complex frequency $z$ by the analytical continuation of
\be
\Pi_{aa'}(q,i\nu_m)=\sum_n\int{dt e^{i(\nu_m-qn)}}\Pi_{aa'}(n,\tau)
\ee
where $\nu_m=2\pi n T$ are bosonic Matsubara frequencies. We denote the analytical continuation of this function to complex frequency $z$ with $\Pi_{aa'}(q,z)$.

\emph{Particle-hole symmetry}  (PHS) requires the Green's function to obey
\be
g_{aa'}(n,\tau)=-(-1)^{a+a'}g_{a'a}(-n,-\tau)
\ee
but plugging this into \pref{eqP} we see that PHS does not impose any restriction on $\Pi(n,\tau)$. In other words, $\Pi(n,\tau)$ is blind to presence or absence of PHS and unlike the strong Kondo regime \cite{Coleman2015}, the spins do not feel the breaking of the particle-hole symmetry.

\emph{Time-reversal symmetry (TRS)} -- In a bi-partite lattice for both electron propagator and susceptibility TRS implies
\be
\hspace{-.2cm}g_{aa'}(n,\tau)=g_{a'a}(-n,\tau),\hspace{2mm} \Pi_{aa'}(n,\tau)=\Pi_{a'a}(-n,\tau). \label{eqTR}
\ee
On the other hand Eq.\,\pref{eqP} generally implies:
\bea
\text{General: }\qquad &\Pi_{aa'}(n,\tau)&=\Pi_{a'a}(-n,-\tau).\label{eqgeneral}
\eea
In absence of TRS, Eq. \pref{eqTR} is violated. However,  due to the relation \,\pref{eqgeneral} which holds generally, the instantaneous $\tau=0$ susceptibility is blind to this TRS breaking. Remarkably the same conclusion holds for the static susceptibility. Transforming these equations to momentum space we find
\bea
\text{TRS: }\qquad&\Pi_{aa'}(q,z)&=\Pi_{a'a}(-q,z), \\
\text{General: }\qquad &\Pi_{aa'}(q,z)&=\Pi_{a'a}(-q,-z). 
\eea
In absence of TRS the first equation is violated. However, again in the static susceptibility limit $z=0$, the equality holds by the second equation. In other words, violation of the TRS is only reflected in the finite-frequency spin susceptibility that is absent in the static RKKY interaction.

\emph{Inversion Symmetry} 
around a center of a link for and the electronic Green's function and the Lindhard function means
\be
\hspace{-.2cm}g_{aa'}(n,\tau)=g_{\bar{a}\bar{a}'}(-n,\tau),\hspace{2mm}\Pi_{aa'}(n,t)=\Pi_{\bar{a}\bar{a}'}(-n,\tau).\label{eqTRS}
\ee
where $\bar A=B$ and $\bar B=A$. Equivalently, in momentum space $\Pi_{aa'}(q,z)=\Pi_{\bar a\bar a'}(-q,z)$. This is a non-trivial restriction even for $\tau=0$ or $z=0$ at least for some components of $\Pi_{aa'}$. This means both static and dynamic RKKY interaction will be affected by the inversion-symmetry of electrons.


To summarize, both instantaneous and static spin susceptibilities $\Pi$ are always time-reversal symmetric even when the TRS is broken in conduction electrons and reflected in $g$. However, dynamic RKKY interaction term is sensitive to the TRS breaking. This property is the central idea of this paper, and is used  to induce TRS-breaking from the electrons to the spins. 

\subsection{Large-N treatment of RKKY interaction}
The problem described by the action \,\pref{eqS} and \pref{eqJeff} is strongly correlated and difficult to deal with. Furthermore, due to the long-range and frustrated coupling induced by the RKKY interaction, the application of Quantum Monte Carlo to this problem suffers from sign-problems \cite{Nefedev2014,Buividovich2017,Nejati2017,Jabar2018,Cahaya2023}. In this section, we apply the large-N approach to this problem. It is important to note that this is different from the standard static \cite{Coleman2015} or dynamic \cite{Komijani2018,Komijani2019,Wang2020,Wang2020b,Shen2020,Han2021,DrouinTouchette2021,Ge2022,Drouin-Touchette2022,Ge2024} large-N treatment of the Kondo problem as the RKKY interaction vanishes in the usual large-N limit.

To this end, we represent the spins using either Abrikosov fermionic $\vec S=f\dg\vec\sigma f$ or Schwinger bosonic $\vec S=b\dg\vec\sigma b$ spinons \cite{Arovas1988}, which results in a quartic interaction, along with a constraint. Next, we generalize the SU(2) group to either SU(N) or SP(N) \cite{Read91,Flint:2008fk,Flint2009hr} groups. This leads to the path integral being dominated by the saddle point (classical) configurations. Usually, a Hubbard-Stratonovitch (HS) transformation is used to decouple the quartic interaction in various channels in terms of quadratic terms. However, following the technology used in Sachdev-Ye-Kitaev model \cite{Sachdev1992,Chowdhury2022}, we perform the entire calculation without a HS transformation. 

A bi-local field $\hat G_f(1,2)$ is introduced via a Lagrange multiplier $\hat \Sigma_f(1,2)$ and the spinons are integrated out to derive an action $S(G_f,\Sigma_f)$ which is extensive in $N$. The saddle point integration at large-N leads to self-consistency equations between Green's function $G$ and self-energy $\Sigma$, which replace the mean-field equation. This method can be applied to all representations of the spin and both SU(N) and SP(N) and is particularly useful in frustrated magnets \cite{Read91}. A detailed derivation of the self-consistency equation in all cases is presented in the Appendix \ref{sec:large}. For brevity, within the paper we focus on the Abrikosov representation
\be
\vec S_{j,\alpha\beta}=f\dg_{j\alpha} f\dn_{j\beta},
\ee
with the constraint
\be
\sum_\alpha f\dg_{j\alpha}f\dn_{j\alpha}=Q,\label{eqQ}
\ee
which is imposed on average using a Lagrange multiplier, $\lambda$, adjusted such that \pref{eqQ} is satisfied. For simplicity, we work with particle-hole symmetric conduction electrons and self-conjugate spins $Q=N/2$. On bipartite lattices with nearest neighbor hopping, the PHS requires $\lambda=0$, but we also consider Kondo triangles for which this does not hold, or Honeycomb lattice with Haldane mass (next nearest-neighbor hoping) for which this holds by a combination of PHS and inversion symmetry. 
Regardless, the $f$-electrons the self-energy is given by [see Appendix\ref{sec:large}]
\be
\Sigma_{ij}(\tau)=-J^{\rm eff}_{ij}(\tau)G_{ij}(\tau).\label{eqself}
\ee
A diagrammatic representation of this self-energy is shown in Fig.\,\ref{fig1}(b), indicating twice scattering of the spinons from the electron gas. This self-energy together with the the Dyson equation for the $f$-electron (spinon) Green's function $G_{ij}(\tau)\equiv\frac{1}{N}\sum_\alpha \sbraket{-T_\tau f_{i\alpha}(\tau)\bar f_{j\alpha}}$,
\be
G^{-1}_{ij}(z)=(z-\lambda_i)\delta_{ij}-\Sigma_{ij}(z)\label{eqDyson}
\ee
with complex frequency $z$, defines a self-consistency problem that is solved numerically here. 

To the best of our knowledge such a simple form \pref{eqself} was not written before. This equation has the important property that it will reinforce $\Sigma_{ij}$ if $J_{ij}$ can be decoupled in an attractive channel (the free energy is concave up) and otherwise, suppress it to zero. Therefore, the formalism automatically takes care of attractive vs. repulsive channels of decoupling the interaction. Solving Eq.\,\pref{eqself} iteratively has the danger that if $G_{ij}=0$ at the first iteration, the self-energy remains zero, even if $J_{ij}\neq 0$. Therefore, we start with a fully random initial configuration and solve Eqs.\,\pref{eqself} and \pref{eqDyson} iteratively to find a self-consistent solution. In practice we solve them in real frequency $z=\omega+i\eta$. 

The self-energy has two parts $\Sigma=\Sigma^{(1)}+\Sigma^{(2)}$  where
\bea
&&\Sigma^{(1)}_{aa'}(k)=\sum_{q} J_H(q)\int \frac{dx}{2\pi} f(x) A^G_{aa'}(k-q,x)\qquad\label{eqself1}
\eea
in terms of momentum-space $J_H(q)=\sum_nJ_H(n)e^{iqn}$ is the frequency-independent Heisenberg part, and 
\bea
&&\Sigma^{(2)}_{aa'}(k,\omega+i\eta)=\frac{J_K^2}{2{\cal N}_s}\sum_q\int{\frac{dxdy}{(2\pi)^2}}\frac{f(x)+n(-y)}{\omega+i\eta-x-y}\nonumber\\
&& \hspace{3.5cm}\times A^\Pi_{aa'}(q,y)A^G_{aa'}(k-q,x)\qquad\label{eqself2}
\eea
is the dynamic RKKY contribution. Both terms are expressed in terms of the spectral function defined as $A_{aa'}^G(k,\omega)\equiv i[G_{aa'}(k,\omega+i\eta)-G_{aa'}(k,\omega-i\eta)]$ and a similar equation for $A^\Pi_{aa'}(k,\omega)$. The first (freq.-independent) part, $\Sigma^{(1)}_{aa'}$ is the mean-field parameters and Eq.\,\pref{eqself1} is essentially the mean-field equation. The static-RKKY \emph{does not} correspond to setting $\omega=0$ in Eq.\,\pref{eqself2}, but rather replacing $J_H(q)$ in Eq.\,\pref{eqself1} with the momentum-space value of the effective Heisenberg interaction \pref{static1}. In the following, we are interested in the $\Sigma^{(2)}(k,z)$, originated from the dynamic RKKY interaction. 


An immediate consequence of the frequency-dependent dynamic RKKY contribution to self-energy \pref{eqself2} is that  the spinon spectral weight can migrate to incoherent parts at high frequencies. However, a full numerical solution to \pref{eqself2} shows that the self-energy has weak frequency dependence. This justifies expanding the $f$-electron self-energy in frequency on two sides of \pref{eqself2}.
\be
\Sigma(k,\omega+i\eta)\approx Z_k^{-1}\sigma(k)+(1-Z_k^{-1})\omega+\dots\label{eqSexpnd}
\ee
resulting in 
\be
G(k,\omega+i\eta)\approx\frac{Z_k}{\omega+i\eta-\sigma(k)}\label{eqG2}
\ee
where $Z_k$ is the spinon's wavefunction renormalization.

We will employ this approximation in the rest of the paper in order to numerically simplify Eqs.\,(\ref{eqDyson}-\ref{eqself2}). We have also solved the problem exactly in the limiting case and found no discrepancies. By a combination of these equations, two coupled equations can be derived for $\sigma_k$ and $Z_k$. Unless specified explicitly, we find that generally $Z_k\approx 1$ and therefore, at least within the models explored here the dynamic-RKKY does not cause major spectral migration to incoherent contributions at high frequencies.

\section{Results}\label{sec:results}
Large-N magnetic systems are prone to dimerization (and trimerization, ...) \cite{Liang1988,Affleck1988}, an instability that is believed to be an artifact of the large-N limit. In the following, we are interested in the translational invariant regime. Therefore, large-N calculations are performed in presence of crystal symmetries to avoid breaking of translational invariance. The details of the symmetry analysis are discussed in Appendix \ref{sec:c3}.\\
\subsection{Spinon lifetime -- localized regime}
At high temperature, the spinons are localized. At the onset of delocalization, we can assume that the momentum-dependent part of the self-energy is small compared with momentum-independent part, i.e. $\lambda+\sigma(k)=\bar\sigma+\delta\sigma(k)$. Expanding the Green's function as $G(k,z)\approx Z(z-\bar\sigma)^{-1}+Z(z-\bar\sigma)^{-2}\delta\sigma(k)$ we allow a complex pole $\bar\sigma=\lambda+i\bar\sigma''$ for the zeroth order. Assuming $q=1/2$ and particle-hole symmetry, the real part of $\bar\sigma$ vanishes $\bar\sigma'=0$, whereas the imaginary part $\bar\sigma''$ can be extracted from $\Sigma_{\rm loc}(\tau)=G_{\rm loc}(\tau)\Pi_{\rm loc}(\tau)$ which translates to a local version of Eq.\,\pref{eqself2}. Using 
\be
A^\Pi(q,y)=\frac{2\pi}{{\cal N}}\sum_{k}\delta(y+\eps_k-\eps_q)[f(\eps_k)-f(\eps_q)]\label{eqAPi}
\ee
and $A^G(x)={2Z\bar\sigma''}/({x^2+\bar\sigma''^2})$ in \pref{eqself2} we find
\be
\bar\sigma''=\bar\sigma''\frac{\tilde J_K^2}{{\cal N}^2}\sum_{kq}   \frac{f(\eps_k-\eps_q)f(-\eps_k)f(\eps_q)}{(\eps_{k}-\eps_{q})^2+\bar{\sigma}''^2}\label{eqlife}
\ee
where $\tilde J_K=ZJ_K$ is the renormalized coupling. For $\bar\sigma''=0^+$, this equation has infrared (IR) divergence at finite temperature. In the wide band (flat density of states) limit and assuming $\rho \tilde J_K\ll 1$ we find $\abs{\bar\sigma''}\approx (\rho\tilde J_K)^2\pi T/4$ is linear in temperature. A similar but slightly longer calculation (see Appendix \ref{sec:lifetime}) shows that $Z=1$ in this localized regime and to this order in Kondo coupling.

Therefore, localized spinons experience a finite lifetime $\tau_{\text{spinon}}\sim{\abs{\bar{\sigma}''}}^{-1}$, originated from the inelastic scattering from the conduction electrons. This can be traced to the retarded nature of the dynamic RKKY, and has to be contrasted with $\abs{\bar\sigma''(T=0)}> 0$ in the strong Kondo coupling regime \cite{Coleman2015}. Such a complex self energy can be interpreted as a finite lifetime of the spinons and is common in the problem of a resonant level coupled to a conduction band. The difference here is that there is no single particle tunneling between spinons and conduction electrons. If the density of state vanishes at the Fermi energy or when conduction electrons are gapped, the inelastic scattering is suppressed more strongly due to the lack of phase space for scattering.

\subsection{Spinon delocalization}
As the temperature is reduced the spinons delocalize due to the magnetic or RKKY interactions. An interesting question is whether the imaginary part of the self-energy, finite at non-zero temperature, affects the onset of the delocalization due to the RKKY interaction. At this temperature, the self-energy can be assumed to be given by the local form, discussed above. {At low temperatures, the delocalization of spinons is expected to result in an energy dispersion. However, as the temperature decreases, solving the saddle point equation for $\sigma_k$ requires increasingly finer resolution, making the problem computationally expensive and challenging. Fortunately, using the Eqs.\,\pref{eqG2} and \pref{eqAPi} the frequency integral in Eq.\,\pref{eqself2} can be done analytically by contour integration technique (see Appendix \ref{sec:lifetime}) and we find that $\sigma_k$ has to satisfy the equation
}
\bw

\bea
\sigma_{k}&=&\sum_{k,q}\frac{(f_k-f_q)\Big[n(\eps_q-\eps_k)+f(\eps_q-\eps_k) \Big]}{(\eps_k-\eps_q)-\sigma_{k-q}}+\frac{1}{\pi}\sum_{k,q}\text{Im}\Big[\frac{f_k-f_q}{\eps_k-\eps_q-\sigma_{k-q}}\Big(\Psi(\frac{1}{2}+\frac{i\sigma_{k-q}}{2\pi T})-\Psi(\frac{1}{2}+i\frac{\eps_k-\eps_q}{2\pi T})\Big) \Big],\qquad\label{eqlong}
\eea

\ew

where $\Psi(x)$ is the digamma function. The second term is purely real, not contributing to $\sigma''_k$. Furthermore, in the local case, $\sigma_k=\bar\sigma$ and \pref{eqlong} reduces back to Eq.\,\pref{eqlife}. 

\begin{figure}[tp]
\includegraphics[width=\linewidth]{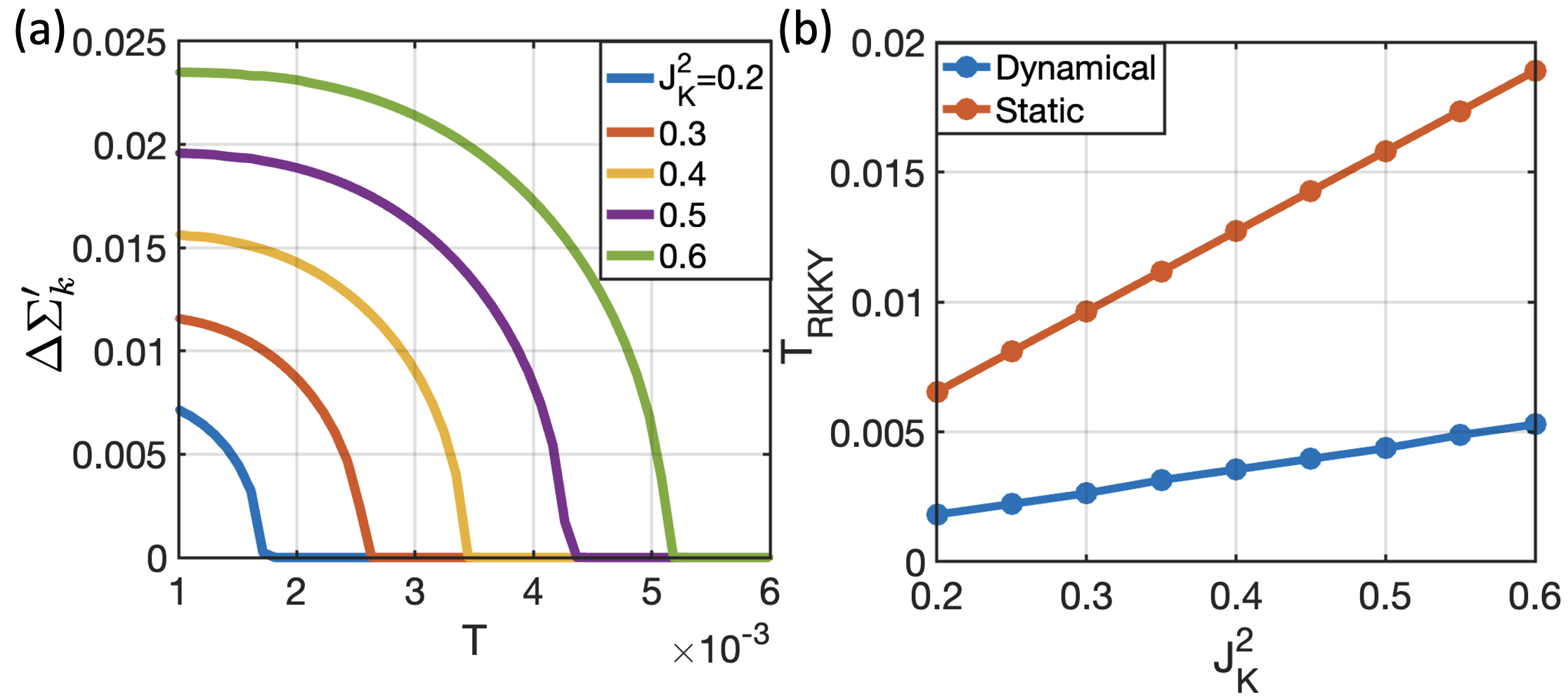}
\caption{Delocalization transitions of spinons in a 1D Kondo lattice in the weak Kondo-coupling regime. (a) The bandwidth of spinons $\Delta\Sigma'_k$ (difference between maximum and minimum value of $\Sigma'_k$) as a function of temperature for various values of $J_K^2$, exhibiting a mean-field behavior transition. All energies are in units of the conduction electron bandwidth $t$. The spinons acquire dispersion below the critical temperature, with larger values of $J_K^2$ leading to higher transition temperatures. (b) Transition temperatures for both dynamic and static RKKY interactions as a function of $J_K^2$. }\label{delocal}
\end{figure}

{The results of solving this equation are illustrated in Fig.\,\ref{delocal}. 
In Fig.\,\ref{delocal}(a), the quantity $\Delta\Sigma'_k$, which represents the difference between the largest and smallest values of the real part of the self-energy $\sigma_k$, is shown. As the temperature increases, $\Delta\Sigma'_k$ decreases steadily, eventually reaching zero at a critical temperature. This marks the point where spinon dispersion vanishes, indicating that the spinons become localized above this temperature. A larger value of $J_K^2$ results in a higher transition temperature for the onset of spinon dispersion. This is because an increase in $J_K^2$ strengthens the RKKY interactions between spinons, thereby stabilizing the delocalized spinon state over a broader range of temperatures. Fig.\,\ref{delocal}(b) shows a linear relation between the transition temperature, $T_{\text{RKKY}}$, and the RKKY interaction strength $J_K^2$ in both static (see Appendix \ref{sec:lifetime}) and dynamic cases, albeit with different slopes. In summary, stronger RKKY interactions allow spinons to remain dispersive at higher temperatures, directly linking the interaction strength to the critical temperature for spinon delocalization.
}

\subsection{RKKY Induced chirality -- Triangle}

\begin{figure}
\includegraphics[width=1\linewidth]{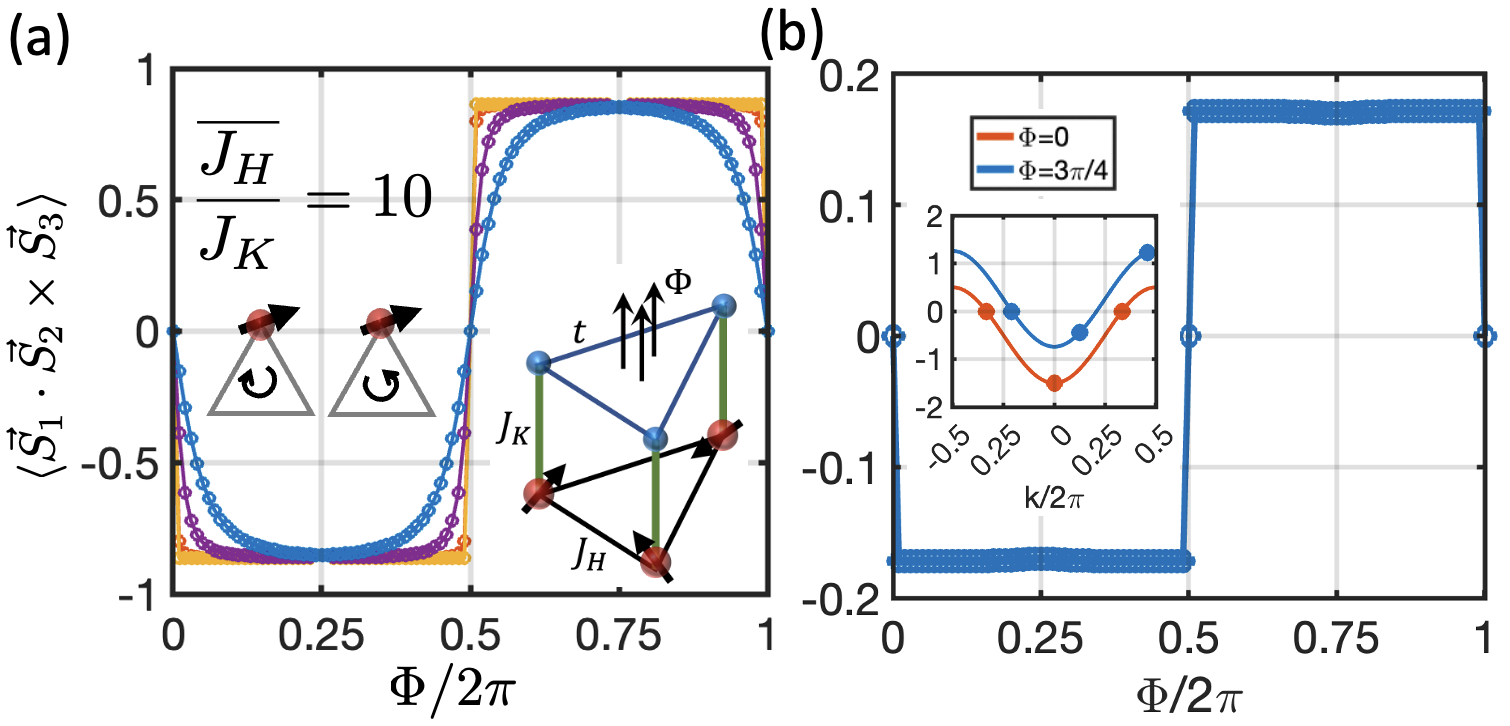}
\caption{The spin chirality $\sbraket{\vec S_1\cdot\vec S_2\times\vec S_3}$ on a Kondo triangle in the weak Kondo regime vs. magnetic flux threading electrons. (a) Exact diagonalization for the spin-1/2 SU(2) problem in presence of various degrees of disorder in the Heisenberg couplings. (b) Large-N results confirm that this induced spin chirality is due to the dynamic RKKY interaction. Insets in (a) shows the setup and the ground state quartet made of two chiral doublets whose degeneracy is lifted by the magnetic field. Inset in (b) shows the spectrum of the triangle (for each spin) as the flux is modified. As $\Phi$ varies, the $\lambda$ shifts the spectrum  in energy to maintain the constraint.}\label{Fig:triangle}
\end{figure}

A non-zero magnetic field coupled to electrons breaks the time reversal symmetry of electrons. Since this is imprinted on the finite frequency parts of $\Pi(k,z)$ we expect it to affect the spins. As a minimal version of the problem, we consider a simple Kondo triangle shown in the inset of Fig.\,\ref{Fig:triangle}(a) in which a magnetic flux $\phi$ threads (only) the conduction electron layer. For simplicity we discard the Zeeman interaction and only consider the impact of the magnetic field on the orbital motion of electrons. 

The advantage of this model is that it is amenable to exact diagonalization, providing a benchmark for the large-N formalism. Fig.\,\ref{Fig:triangle}(a) shows the spin chirality of the spin-1/2 SU(2) version of the model, defined as \cite{wen2004} 
\bea
\chi&\equiv&\langle S_1\cdot S_2\times S_3 \rangle = i\sum_{\alpha\beta\gamma}S_{1,\alpha\beta}S_{2,\beta\gamma}S_{3,\gamma\alpha}\nonumber
\eea
as a function of the flux $\phi$, computed using exact diagonalization. At TRS-preserving fluxes $\phi=0,\pi$, the ground state is a quartet, composed of chiral and anti-chiral spin-doublets. The flux lifts the chirality degeneracy resulting in the net chirality. The chirality is robust against introduction of weak disorder in the $J^H_{ij}$ Heisenberg couplings, although the steps become smooth.

Fig.\,\ref{Fig:triangle}(b) shows the same quantity computed in the large-N limit from the self-consistent solution to Eqs.\,\pref{eqself} and \pref{eqDyson}. In this case, we can write
\bean
\chi&\to&\sum_{\alpha\beta\gamma}\frac{8}{N^3}\text{Im}\langle f_{1\alpha}^\dagger f_{1\beta} f_{2\beta}^\dagger f_{2\gamma} f_{3\gamma}^\dagger f_{3\alpha} \rangle\\
&&\hspace{1cm}=8\,\text{Im}\,[G_{21}(0^-)G_{32}(0^-)G_{13}(0^-)]\\
&&\hspace{3cm}=\text{Im}\,[\frac{2}{3}\sum_k e^{ik}G_{k}(0^-)]^3.
\eean

As seen in Fig.\,\ref{Fig:triangle}(b), the main feature of the SU(2) model is reproduced in the large-N limit, albeit the amplitude of the signal is weaker. Using our simplifying assumption that the self-energy, even in presence of the dynamic RKKY, is just a frequency-independent dispersion $\Sigma(k,\omega)=\sigma_k$, the magnitude of the chirality can be computed. In this case, the lesser Green's function is just given by $G_k(0^-)=f(\sigma_k)$, where $f(x)\equiv 1/(e^{\beta x}+1)$ is the Fermi-Dirac distribution.

On the electron layer, the magnetic flux $\phi$ effectively modifies the electronic dispersion in $g_k(z)=(z-\eps_k)^{-1}$ from $\eps_k=-2t\cos k$ to $\eps_{k+\phi/3}$. With the $f$-electron also we choose a gauge in which $k=2n\pi/3$. Denoting
\ben
f_0=f(\sigma_0),\hspace{2mm} f_1=f(\sigma_{\frac{2\pi}{3}}),\hspace{2mm} f_{-1}=f(\sigma_{\frac{-2\pi}{3}})
\een
the chirality becomes
\ben
\chi=\text{Im}[\frac{2}{3}(f_0+e^{i \frac{2\pi}{3}}f_1+e^{-i \frac{2\pi}{3}}f_{-1})^3].
\een
The constraint here requires that $f_0+f_1+f_{-1}=3/2$.  The precise values of $\sigma_k$ require a solving Eq.\,\pref{eqlong}. However, the following arguments can be generally made.

We always have $f_0=1$ as this state is fully occupied. In the TRS-preserving case $\phi=0,\pi$, we have $f_1=f_{-1}=1/4$ such that $\chi=0$. The maximum of $\chi=\text{Im}[\frac{2}{3}(1+\frac{1}{2}e^{i\frac{\pm 2\pi}{3}})]^3=\pm 0.1925$ when $f_1=1/2, f_{-1}=0$ or $f_1=0, f_{-1}=1/2$, in agreement with Fig.\,\ref{Fig:triangle}(b). A simple way for this to happen is that f-electrons still obey $\sigma_k=\lambda-2t_f\cos\ k$, but their flux $\phi_f$ is locked to $\phi$ so that $k_f=2n\pi/3+\phi$, as shown in the inset of Fig.\,\pref{Fig:triangle}.

It is noteworthy that our analysis was done in the weak Kondo coupling regime $J^K\ll J^H, T$. In the opposite regime of strong Kondo coupling, this effect \cite{Feng2013} is known and understood \cite{Wang2024}. In that case, the non-zero emergent hybridization between $c$ and $f$ electrons results in Kondo flux repulsion and the internal gauge field of $f$ electrons is locked to the external gauge field affecting $c$ electrons. The new feature here, is that this results holds in the weak perturbative Kondo coupling regime as well.

\begin{figure}[h!]
\includegraphics[width=.9\linewidth]{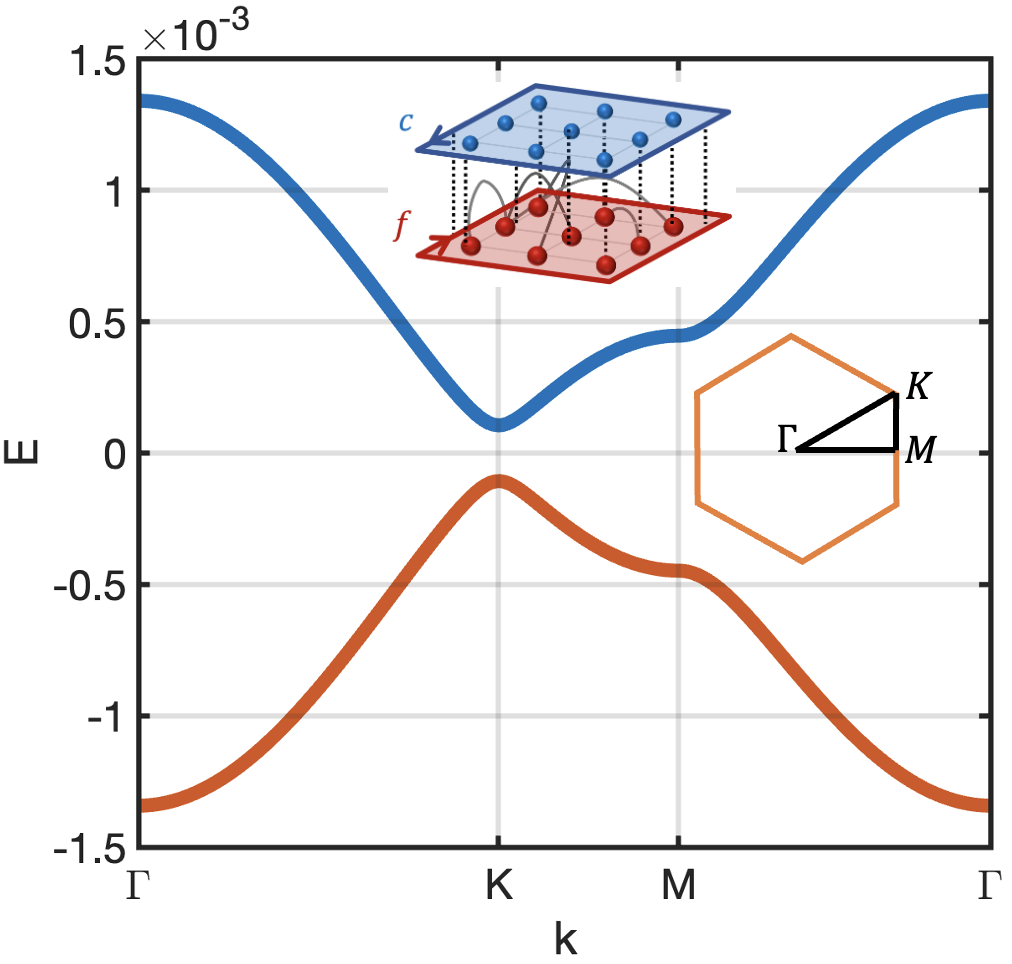}
\caption{The spinon dispersion due to the dynamic RKKY interaction in a Kondo lattice with honeycomb geometry. The cut through the Brillouin zone is indicated in the inset. The inversion symmetry was preserved throughout the calculation. The time-reversal symmetry breaking in the conduction layer, induced by Haldane \cite{Haldane1988} fluxes, is transferred by the dynamic RKKY effect to the spinons, opening a gap at the Dirac cones. The parameters of the model are $J_K/t_1=0.3$ and $t_2/t_1=0.8$ where $t_1$ and $t_2$ are the magnitudes of nearest neighbor and next-nearest neighbor hopings. As indicated in the inset, with open boundary conditions, the edge mode in the c-layer is mirrored by the $f$-layer.}\label{haldane}
\end{figure}

It is also illuminating to consider the effect of the discarded Zeeman splitting on this physics. Since the ground state is composed of a doublet (or a multiplet) of chiral states in the SU(2) case (in the SU(N) case), lifting the spin degeneracy can reduce the overall chirality by populating the occupied states with more spinons of minority chirality. But in both cases, the Zeeman splitting has to overcome a threshold to reduce the spin chirality.

\subsection{RKKY Induced chirality -- Haldane Model}
The Kondo-Heisenberg model on the honeycomb lattice has been studied in the past \cite{saremi2007,Feng2013}, as a fascinating platform for uncovering new physics, including topological order in the multi-channel case \cite{Ge2024}. In this section, we take the advantage of the dynamic RKKY interaction in the weak Kondo limit to induce TRS breaking from conduction electrons to the spin-liquid on a honeycomb lattice \cite{Haldane1988}. This leads to a gap opening in the spin-liquid which stabilizes chiral edge states.

On a bi-partite lattice, the Green's function of electrons is $\bb g_k(z)=[z-\bb H_{k}]^{-1}$, where $\bb g_k(z)$ and $\bb H_{k}$ are two dimensional matrices on the sub-lattice space. The Green's function can be written as 
\be
\bb g_k(z)=\frac{1}{2}\Big[\frac{\bb 1+\bb N(k)}{z-\eps_k}+\frac{\bb 1-\bb N(k)}{z+\eps_k}\Big], 
\ee
where the explicit form of $\bb N(k)$ and $\eps(k)$ will depend on the specific model. Assuming PHS and inversion symmetry, $\bb\Sigma(k,z)=-\tau^y\bb\Sigma^T(-k,-z)\tau^y$. Expanding both sides of this equation using \pref{eqSexpnd} we find that $\bb Z_k=Z_k\bb 1$ is a scalar. Consequently,  the Green's function of spinons $\bb G_f(z,k)\equiv Z_k[z-\bb \sigma(k)]^{-1}$ can be expressed as
\ben
\bb G_f(k,z)=\frac{1}{2}\Big[\frac{1+\bb M(k)}{z-E_k}+\frac{1-\bb M(k)}{z+E_k} \Big]Z_k
\een
where $ \sigma(k)\equiv\vec \sigma(k)\cdot\vec\tau$ with $\vec\tau$ Pali matrices acting in the A-B sublattice, where $\vec\sigma(k)$ are the coefficients of expansion of the 2$\times$2 matrix $\sigma(k)$ in terms of the Pauli matrices. Using $\sigma_+(k)=\sigma_x(k)+i\sigma_y(k)$, we can write
\ben
\bb M=\frac{1}{E_k}\mat{
\sigma_3&\sigma_-\\
\sigma_+&-\sigma_3}, \qquad E_k=\sqrt{\sigma^2_3+\abs{\sigma_+}^2}
\een

We have found that again $Z_k\approx 1$ throughout the Brillouin zone in this problem.   Therefore, with the help of the $g_k(z)$ and $G_k(z)$ and Eq\,\pref{eqself2}, the saddle point equation for self-energy of spinons, can be rewritten as (see Appendix \ref{sec:haldane} for the derivation)
\bw
\small{
\be
\Re\sigma^{(2)}_{aa'}(k)=-\frac{J_K^2}{8{\cal N}_s^2}\sum_{pq}\Big\{\frac{1}{E_{k-q}+\eps_{p^+}+\eps_{p^-}}\Big[M_{aa'}(k-q)\Big(\delta_{aa'}-N_{aa'}(p^+)N_{a'a}(p^-)\Big)+\delta_{aa'}\Big(N_{aa'}(p^+)-N_{a'a}(p^-)\Big)\Big]\Big\}\label{eq8}
\ee}

\ew

where $p^\pm\equiv p\pm q/2$. Since we have imposed PHS and inversion symmetries, any diagonal contribution is odd-in-$k$, breaking TRS and opening a gap in the spectrum.  It is simple to see that the last term on the right hand side, invalidates a possible $\sigma_{aa}=0$ solution, potentially opening a gap. Assuming lattice inversion symmetry, the only origin of the gap is the TRS breaking symmetry.

Fig\,\pref{haldane} shows the result of a full numerical calculation of the self-consistency equation. The eigenvalues of the spinon's self-energy exhibit a gap $K$ point around the reduced Brillouin zone. Therefore, we have shown that the spinons mirror the topology (and thus the edge states) of the conduction electrons and this can be captured in the weak Kondo-coupling limit by  the dynamic-RKKY.

\section{Conclusion}\label{sec:conclusion}
In this paper we have shown that a weak Kondo interaction between spins and conduction electrons produces a dynamic RKKY interaction which is often approximated as the static RKKY interaction. We have used symmetry analysis to argue that this approximation fails to capture some crucial physics, including time-reversal symmetry breaking. To the best of our knowledge this was never pointed out before. We have used large-N theory to analyze the dynamic RKKY interaction using both SU(N) and SP(N) spins with either Schwinger boson or Abrikosov fermions (Appendix \ref{sec:large}). In particular, for SU(N) spins with Abrikosov fermions we have shown that due to dynamic RKKY interaction, spinons have a finite lifetime and wavefunction renormalization, although the latter was found to remain unity within the models considered here. We have compared the spinon delocalization temperature between dynamic and static RKKY interactions. Furthermore, it is possible to induce chirality into the spin liquid by using either magnetic field or proximity to a topological insulator.  This proof of principles opens a window to induce spin chirality and topology on spin systems using proximity to topologically non-trivial electronic systems. Therefore, dynamic RKKY can act as a tunable Dzyaloshinskii-Moriya interaction even in centrosymmetric systems. 

This result also helps to understand the recently discovered topological order in weakly Kondo-coupled multi-channel Kondo lattices \cite{Ge2024}. The dynamic RKKY induced topology on the f-electrons leads to an anti-chiral edge states in the spin-layer. This edge state is however Kondo-coupled to the multiple chiral edge states of conduction electrons, resulting in partial gapping and a fractional edge state, which by the bulk-boundary correspondence, indicates topological order in the bulk. In the single-channel case, this chiral-antichiral coupling leads to a complete gapping of spin degrees of freedom at sufficiently low temperatures, leaving only the gapless charge mode behind. Therefore, the physics proposed here holds for low but finite temperatures. 


Our results can be extended in a number of ways. In this work, we have focused on using electrons to tune spinons, however, the reverse process is also possible. At weak Kondo coupling, the electron states can provide insight into the spin configuration. For instance, in the presence of a chiral spin liquid, electrons experience an effective orbital magnetic field. More generally, the second-order in Kondo coupling expansion, also provides a way to read off dynamic spin susceptibility $\chi_f(q,z)$ of $f$-electrons by studying the optical conductivity of $c$-electrons. This is useful for when a direct measurement of $\chi_f(q,z)$ using spectroscopic tools is not possible, in line with recent theoretical proposals \cite{Mazzilli2023}, but under equilibrium conditions.  It would be interesting to show that the topology-inducing mechanism discovered here also works for arbitrary microscopic model, i.e.\,the $Z_2$ topological index of the $f$-electrons is related to that of $c$-electrons. Finally, RKKY interaction may help to avoid spin dimerization. However, such applications require a study of 1/N effects to avoid the spurious dimerization induced by the large-N effects. We have not explored these directions in the present work and leave it for future investigations.


\appendix
\section*{Appendix}
\subsection{Dynamic RKKY interaction}\label{sec:app0}
The Hamiltonian has three parts as
\bean
H_0&=&\sum_k \eps_k c^\dagger_k c_k\\
H_H&=&\sum_{ij}J_H S_{i,\alpha\beta}S_{j,\beta\alpha}\\
H_K&=&\sum_j  J_K S_{j\alpha\beta}c_{j\beta}^{\dagger}c_{j\alpha},
\eean
where $H_0$ is the free electron part, $H_H$ is the Heisenberg interaction and $H_K$ is Kondo interaction. Assume $H_K$ is small compared with $H_0+H_H$ denoted as $\tilde{H}_0$.

\ben
H=\underbrace{H_0+H_H}_{\tilde{H}_0}+H_K.
\een
In time-ordered perturbation theory, we have  
\ben
e^{-\beta H}=e^{-\beta \bar{H}_0}\underbrace{e^{\beta \tilde{H}_0} e^{-\beta H}}_{S(\beta)}
\een
where
\ben
S=T_\tau e^{-\int_0^\beta d\tau V_I(\tau)}.
\een
Here the $V_I=H_K$ as $\tilde{H}_0=H_0+H_H$ such that the S-matrix is
\ben
S=T_\tau \text{exp} \Big[ -\int_0^\beta d\tau H_K(\tau) \Big].
\een
Then one can expand the interaction part to the second order and recover it back to the exponential form as following:
\bean
&&e^{-\int_0^{\beta} d\tau H_K(\tau)}\\
&\approx&1+\frac{J_K^2}{2}\int_0^{\beta} d\tau_1 d\tau_2 H_K(\tau_1)H_K(\tau_2)\\
&=&1-\frac{J_K^2}{2}\int_0^{\beta} d\tau_1 d\tau_2 S_{i,\alpha\beta}(\tau_1)g_{ij}(\tau_1,\tau_2)g_{ji}(\tau_2,\tau_1)S_{j,\beta\alpha}(\tau_2)\\
&=&\text{exp}\left[-\frac{J_K^2}{2}\int_0^{\beta} d\tau_1 d\tau_2 S_{i,\alpha\beta}(\tau_1)\Pi_{ij}(\tau_1,\tau_2)S_{j,\beta\alpha}(\tau_2)\right].
\eean
Now, the action can be written in the path integral as
\ben
S=S_B+\intob{d\tau d\tau'}S_{i,\alpha \beta}(\tau)J_{ij}^{\text{eff}}(\tau,\tau')S_{j, \beta\alpha}(\tau')
\een
where in absence of spin-orbit and electron-electron interactions
\ben
J^{\text{eff}}_{ij}(\tau-\tau')=J^H_{ij}\delta(\tau-\tau')+\frac{J_K^2}{2}\Pi_{ij}(\tau-\tau'),
\een
and
\ben
\Pi_{ij}(\tau-\tau')=g_{ij}(\tau-\tau')g_{ji}(\tau'-\tau),
\een
and the $S_B$ is Berry term of the action.
\begin{figure}
\centering
\includegraphics[scale=0.5]{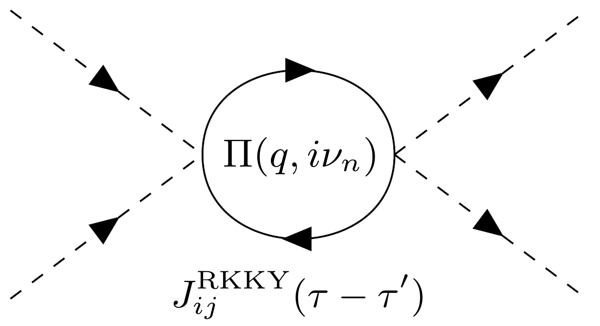}
\caption{Feynman diagram of dynamic RKKY interaction}
\end{figure}


\subsection{Symmetry Analysis}\label{sec:app1}
\emph{Inversion Symmetry} -- Here, we consider lattices that have inversion symmetry with respect to the center of a link. The model may or may not have inversion symmetry, depending on the Hamiltonian. This includes both linear and honeycomb lattices considered in the paper. On a two-atom per unit cell lattice, the inversion reflects the lattice $\vec n\to-\vec n$ and exchanges sublattices $A\to \bar A=B$ and $B\to \bar B=A$. In presence of the symmetry, $[{\cal I},H]=0$ from which we can write
\bean
\braket{{\cal I}^{-1}{\cal I}A_a(n,t),B_{a'}}
&=&\braket{A_{\bar a}(-n,t)B_{\bar{a'}}}
\eean
From this we conclude
\bean
\Pi^{\rm ret}_{aa'}(n,t)&=&-i\theta(t)\sum_i\braket{({\cal I}^{-1}{\cal I})[O^i_a(n,t),O^i_{a'}(0,0)]}\nonumber\\
&=&\sum_i\braket{-i\theta(t)[O^i_{\bar a}(-n,t),O^i_{\bar{a'}}(0,0)]}
\eean
The same holds in imaginary time:
\ben
\Pi_{aa'}(n,\tau)=\Pi_{\bar a \bar {a}'}(-n,\tau).
\een
In momentum $k$ and complex frequency $z$ space:
\ben
\Pi_{aa'}(k,z)=\Pi_{\bar a \bar a'}(-k,z).
\een

\emph{Time-reversal Symmetry} -- With time-reversal symmetry we have the choice whether ${\cal T}^2=\pm 1$. Since we work with SU(N) spins and orbital effects of magnetic field we could use either definitions. The following holds in both cases. Here, it's best to pick a representation so that states and operators behave as matrices acting in the many-body space. Assuming $[{\cal T},H]=0$, an arbitrary real-time correlation function can be written as
\bean
\braket{A(t)B\dg}&=&\braket{{\cal T}^{-1}{\cal T}A(t)B\dg}\\
&=&\braket{{\cal T}^{-1}A^*(-t)(B\dg)^*{\cal T}}\\
&=&Z^{-1}\sum_n\braket{n^*\vert A^*(-t)(B\dg)^* \vert n^*}\\
&=&Z^{-1}\sum_n\braket{n\vert B A\dg(-t) \vert n}\\
&=&\braket{B(t) A\dg}
\eean
In the third line we have used that $\ket{n^*}={\cal T}\ket{n}$ and ${\cal T}A{\cal T}^{-1}=A^*$. Operator expressions like $A^*$ or $A^T$ are representation-dependent but the final expression is not. In the fourth line we have used the relation \cite{Gurarie2011}
\ben
\braket{n^*\vert A\vert m^*}=\braket{m\vert A^T \vert n}.
\een
This relation can be used to state the symmetry property of the retarded function. However, the same statement holds in imaginary time, assuming that ${\cal T}: \tau\to -\tau$ under time-reversal:
\ben
\braket{-T_\tau A(\tau)B\dg}=\braket{-T_\tau B(\tau)A\dg},
\een
leading to Eqs.\,\pref{eqTRS}. In momentum space this means
\ben
\hspace{-.5cm}g_{aa'}(k,z)=g_{a'a}(-k,z), \quad \Pi_{aa'}(k,z)=\Pi_{a'a}(-k,z),
\een
for the complex frequency $z$.

\emph{Particle-Hole symmetry} -- On a bi-partite lattice we assume
\ben
{\cal C} c\dn_{na} {\cal C}^{-1}=(-1)^ac\dg_{na}.
\een
where $(-1)^{A/B}=\pm 1$. Assuming $[{\cal C},H]=0$, we have
\bean
g_{aa'}(n,\tau)&=&\sbraket{-{\cal C}^{-1}{\cal C} T_\tau c\dn_{na}(\tau)c\dg_{0a'}}\nonumber\\
&=&-(-1)^{a+a'}\sbraket{- T_\tau c\dn_{0a'}(-\tau)c\dg_{na}}\nonumber\\
&=&-(-1)^{a+a'}g_{a'a}(-n,-\tau).
\eean

\emph{$C_3$ symmetry}\label{sec:c3} - The self-energy in the Haldane case has $C_3$ symmetry which means unit vectors rotate $2\pi/3$ angle doesn't change the system. The rotation matrix of $\theta={2\pi}/{3}$ for counterclockwise is
\bean
R(\theta)&=&\left[\begin{array}{cc} \cos(\theta) & -\sin(\theta)\\ \sin(\theta)& \cos(\theta)\\ \end{array}\right]=\left[\begin{array}{cc} -\frac{1}{2} & -\frac{\sqrt{3}}{2}\\ \frac{\sqrt{3}}{2}& -\frac{1}{2}\\ \end{array}\right]\\
\eean
when we rotates the $a_1,a_2$ (real space vectors) with $\frac{2\pi}{3}$ in counterclockwise, the $b_1,b_2$ in reciprocal space rotates in the same direction. Here we define the moment 
\ben
k=\frac{n}{N}b_1+\frac{m}{N}b_2.
\een
After rotation
\ben
k=\frac{n}{N}b'_1+\frac{m}{N}b'_2=\frac{n}{N}(-b_1-b_2)+\frac{m}{N}b_1=(\frac{m-n}{N})b_1-\frac{n}{N}b_2
\een
which means the rotation of $k$ can be expressed as
\ben
\left[\begin{array}{c} n'\\m' \end{array}\right]=\left[\begin{array}{cc} -1&1\\-1 &0\\ \end{array}\right]\left[\begin{array}{c} n\\m \end{array}\right].
\een
For the $c_{B,k}$
\bean
c_{B,k}=\frac{1}{\sqrt{N}}\sum_{i} e^{ikR_i} c_{B,i}.
\eean
After rotation
\ben
c_{B,k'}=\frac{1}{\sqrt{N}}\sum_{i} e^{ik'R_i} c_{B,i}=c_{B,k}e^{ika_1}.
\een
This gives the rotation matrix $R$ for the momentum $k$
\ben
R=\left[\begin{array}{cc}
-1&1\\-1&0
\end{array}\right],
\een
and at the same time, the unitary transformation of the self-energy is 
\ben
U=\left [ \begin{array}{cc} 1& \\ & e^{ika_1}\\ \end{array}\right ],
\een
such that after the $C_3$ rotation the self-energy becomes
\ben
\sigma(k)=U^\dagger \sigma(Rk) U.
\een

\begin{figure}
\centering
\includegraphics[scale=0.5]{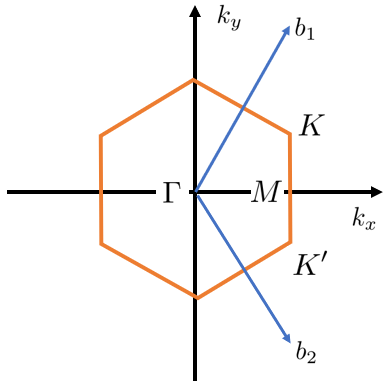}
\caption{Brillouin zone of honeycomb lattice}
\end{figure}

\subsection{Large-N treatment}\label{sec:large}
We first discuss SP(N) representations using both Schwinger bosons and Abrikosov fermions. The SU(N) case, can be obtained from that by discarding some of the terms. Although the full formalism is not used in the paper, this technology is useful for treating (possibly dynamic) frustrated magnetic interaction between spins.

\subsubsection{Sp(N) Schwinger bosons}
In the $Sp(N)$ case of Schwinger bosons, the spin operator will be
\ben
S_{i,\alpha\beta}=b^\dagger_{i,\alpha}b_{i,\beta}-\tilde{\alpha}\tilde{\beta}b^\dagger_{i,-\beta}b_{i,-\alpha}
\een
where $b^\dagger$ and $b$ are bosonic operators. Here we define four different Green's functions (particle-particle, hole-hole, particle-hole and hole-particle)
\bean
G^{pp}_{ij}(\tau,\tau')&=&-\frac{1}{N}T_{\tau}\langle b_{i,\alpha}(\tau) b^\dagger_{j,\alpha}(\tau')\rangle\\
G^{hh}_{ij}(\tau,\tau')&=&-\frac{1}{N}T_{\tau}\langle b^\dagger_{i,-\alpha}(\tau) b_{j,-\alpha}(\tau')\rangle\\
G^{ph}_{ij}(\tau,\tau')&=&-\frac{1}{N}T_{\tau}\langle \tilde{\alpha} b_{i,\alpha}(\tau) b_{j,-\alpha}(\tau')\rangle\\
G^{hp}_{ij}(\tau,\tau')&=&-\frac{1}{N}T_{\tau}\langle \tilde{\alpha} b^\dagger_{i,-\alpha}(\tau) b^\dagger_{j,\alpha}(\tau')\rangle.
\eean
After plugging these Green's functions to the action,  the action becomes
\bw

\ben
S=\int d\tau d\tau' \sum_{i,\alpha}\bar{c}_{i,\alpha}\partial_{\tau}c_{i,\alpha}+\sum_{i,\alpha}\bar{f}_{i,\alpha}\partial_{\tau}f_{i,\alpha}-J_{ij}^{\text{eff}}\eta^{\mu\mu'}(\tau,\tau')G^{\mu\mu'}_{ij}(\tau,\tau')G^{\mu'\mu}_{ji}(\tau',\tau)-\Sigma^{\mu\mu'}_{ij}(\tau,\tau')\left[G^{\mu'\mu}_{ji}(\tau',\tau)+\frac{1}{N} \bb B_{ji}^{\mu'\mu} \right]
\een

\ew

where $\mu\mu'=pp,ph,hp,hh$ and $\eta^{\mu\mu'}=1$ for $\mu\mu'=pp,hh$ while $\eta^{\mu\mu'}=-1$ for $\mu\mu'=ph,hp$. The matrix $\bb B_{ji}$ has the form as
\ben
\bb B_{ji}=\left [\begin{array}{cc}
-b_{j,\alpha}(\tau') b^\dagger_{i,\alpha}(\tau)&\tilde{\alpha}b_{j,\alpha}(\tau') b_{i,-\alpha}(\tau)\\\tilde{\alpha}b^\dagger_{j,-\alpha}(\tau') b^\dagger_{i,\alpha}(\tau)&-b^\dagger_{j,-\alpha}(\tau') b_{i,-\alpha}(\tau)
\end{array}\right].
\een
From the condition $\delta S/\delta G=0$ in large-N theory, saddle point equations are obtained
\ben
\Sigma^{\mu\mu'}_{ij}(\tau,\tau')=-J^{\text{eff}}_{ij}(\tau,\tau')G^{\mu\mu'}_{ij}(\tau,\tau').
\een

\subsubsection{Sp(N) Abrikosov fermions}
In the $Sp(N)$ case of Abrikosov fermions,
\ben
S_{i,\alpha\beta}=f^\dagger_{i,\alpha}f_{i,\beta}-\tilde{\alpha}\tilde{\beta}f^\dagger_{i,-\beta}f_{i,-\alpha}
\een
where $f^\dagger$ and $f$ are fermionic operators. Similarly, we define four different Green's functions (particle-particle, hole-hole, particle-hole and hole-particle)
\ben
G^{pp}_{ij}(\tau,\tau')=-\frac{1}{N}T_{\tau}\langle f_{i,\alpha}(\tau) f^\dagger_{j,\alpha}(\tau')\rangle
\een

\ben
G^{hh}_{ij}(\tau,\tau')=-\frac{1}{N}T_{\tau}\langle f^\dagger_{i,-\alpha}(\tau) f_{j,-\alpha}(\tau')\rangle
\een

\ben
G^{ph}_{ij}(\tau,\tau')=-\frac{1}{N}T_{\tau}\langle \tilde{\alpha} f_{i,\alpha}(\tau) f_{j,-\alpha}(\tau')\rangle
\een

\ben
G^{hp}_{ij}(\tau,\tau')=-\frac{1}{N}T_{\tau}\langle \tilde{\alpha} f^\dagger_{i,-\alpha}(\tau) f^\dagger_{j,\alpha}(\tau')\rangle.
\een
After plugging these Green's functions to the action, the action becomes
\bw

\ben
S=\int d\tau d\tau' \sum_{i,\alpha}\bar{c}_{i,\alpha}\partial_{\tau}c_{i,\alpha}+\sum_{i,\alpha}\bar{f}_{i,\alpha}\partial_{\tau}f_{i,\alpha}-J_{ij}^{\text{eff}}(\tau,\tau')G^{\mu\mu'}_{ij}(\tau,\tau')G^{\mu'\mu}_{ji}(\tau',\tau)-\Sigma^{\mu\mu'}_{ij}(\tau,\tau')\left[G^{\mu'\mu}_{ji}(\tau',\tau)+\frac{1}{N} \bb F_{ji}^{\mu'\mu} \right]
\een

\ew

where $\mu\mu'=pp,ph,hp,hh$ and 
\ben
\bb F_{ji}=\left [\begin{array}{cc}
f_{j,\alpha}(\tau') f^\dagger_{i,\alpha}(\tau)&\tilde{\alpha}f_{j,\alpha}(\tau') f_{i,-\alpha}(\tau)\\\tilde{\alpha}f^\dagger_{j,-\alpha}(\tau') f^\dagger_{i,\alpha}(\tau)&f^\dagger_{j,-\alpha}(\tau') f_{i,-\alpha}(\tau)
\end{array}\right].
\een
From the condition $\delta S/\delta G=0$ in large-N theory, saddle point equations are
\ben
\Sigma^{\mu\mu'}_{ij}(\tau,\tau')=-J^{\text{eff}}_{ij}(\tau,\tau')G^{\mu\mu'}_{ij}(\tau,\tau').
\een

\subsubsection{SU(N) Abrikosov fermions}
For $SU(N)$ case where $S_{i,\alpha\beta}=f_{i\alpha}^\dagger f_{i\beta}$, only $G^{pp}$ and $G^{hh}$ are included. The saddle point equations in the time and real space are
\bean
\Sigma_{ij}(\tau,\tau')&=&-J_{ij}^{\text{eff}}(\tau,\tau')G_{ij}(\tau,\tau')\\
(\bb G^{-1})_{ij}(\tau,\tau')&=&(\partial_{\tau}+\lambda)\delta_{ij}\delta(\tau,\tau')+\Sigma_{ji}(\tau',\tau)\\
q&=&\frac{1}{N}\sum_i \langle f_i^\dagger f_i \rangle.
\eean
In the main text, we focus on this SU(N) case.

\subsection{Self-Energy in Spectral Function Form}\label{sec:spec}
In this section, we will derive Eq.\,\pref{eqself2} from Eq.\,\pref{eqself}. Assume translational invariance both for time and space, the self-energy $\Sigma_{ij}(\tau,\tau')\propto \Sigma_{i-j}(\tau-\tau')$. Start from
the first part of the self energy 
\bean
\Sigma^{(1)}_{aa'}(i-j)=-J_H G_{aa'}(i-j,\tau-\tau')\delta(\tau,\tau').
\eean
After Fourier transform to the momentum space and Matsubara space, it becomes
\bean
\Sigma^{(1)}_{aa'}(k)&=&-\frac{1}{\beta}\sum_{q,n} J_H(q) G_{aa'}(k-q,i\omega_n)\\
&=&\sum_{q} J_H(q) \oint \frac{dz}{2\pi i}f(z)G_{aa'}(k-q,z)\\
&=&\sum_{q} J_H(q) \oint \frac{dz}{2\pi i}f(z)\int \frac{dx}{2\pi i}\frac{A^G_{aa'}(k-q,x)}{z-x}\\
&=&\sum_{q} J_H(q)\int \frac{dx}{2\pi i} f(x) A^G_{aa'}(k-q,x)\\
\eean
where $A_{aa'}^G(x)$ is the spectral function of spinon's Green's function $G_{aa'}$. The second part of the self-energy is
\ben
\Sigma^{(2)}_{aa'}(i-j,\tau-\tau')=-\frac{1}{2}J_K^2 \Pi_{aa'}(i-j,\tau-\tau')G_{aa'}(i-j,\tau-\tau')\\
\een
where the $\Pi_{aa'}$ is
\ben
\Pi_{aa'}(i-j,\tau-\tau')=g_{aa'}(i-j,\tau-\tau')g_{a'a}(j-i,\tau'-\tau)
\een
In momentum and frequency spaces we have
\ben
\Sigma^{(2)}_{aa'}(k,i\omega_m)=-\frac{J^2_K}{2\beta N}\sum_{n,q}\Pi_{aa'}(q,i\nu_n)G_{aa'}(i\omega_m-i\nu_n,k-q)
\een
and
\ben
\Pi_{aa'}(q,i\nu_n)=\frac{1}{\beta N}\sum_{m,k}g_{aa'}(k+q,i\nu_n+i\omega_m)g_{a'a}(k,i\omega_m)
\een
Green's function matrix $g_{aa'}$ in the Lindhard function $\Pi_{aa'}$ is
\ben
g_{aa'}(k,z)=\frac{P_{aa'}^+(k)}{z-\eps_k}+\frac{P_{aa'}^-(k)}{z+\eps_k}, \hspace{2mm} P^{\pm}(k)=\frac{1\pm N_{aa'}(k)}{2}
\een
where $N(k)$ for different lattices has different forms.
\bw

The Lindhard function can be shown by using spectral function as
\bean
\Pi_{aa'}(q,i\nu_n)=\frac{1}{N}\sum_{k}\Big[&&P^+_{k+q,aa'}P^+_{k,a'a}\frac{f(\eps_k)-f(\eps_{k+q})}{i\nu_n+\eps_k-\eps_{k+q}}+P^+_{k+q,aa'}P^-_{k,a'a}\frac{f(-\eps_k)-f(\eps_{k+q})}{i\nu_n-\eps_k-\eps_{k+q}}\\
+&&P^-_{k+q,aa'}P^+_{k,a'a}\frac{f(\eps_k)-f(-\eps_{k+q})}{i\nu_n+\eps_k+\eps_{k+q}}+P^-_{k+q,aa'}P^-_{k,a'a}\frac{f(-\eps_k)-f(-\eps_{k+q})}{i\nu_n-\eps_k+\eps_{k+q}}\Big],
\eean
and the self-energy is
\bean
\Sigma^{(2)}_{aa'}(k,i\omega_m)&=&-\frac{J^2_K}{2\beta N}\sum_{n,q}\Pi_{aa'}(q,i\nu_n)G_{aa'}(k-q,i\omega_m-i\nu_n)\\
&=&-\frac{J^2_K}{2N}\sum_{q}\oint \frac{dz}{2\pi i}n_B(z)\frac{dx dy}{(2\pi)^2}\frac{A_\Pi^{aa'}(q,x)}{z-x}\frac{A_G^{aa'}(k-q,y)}{i\omega_m-z-y}\\
&=&\frac{J_K^2}{2N}\sum_q \int \frac{dx}{2\pi} \Big[ n_B(x)A_{\Pi}^{aa'}(q,x)G_{aa'}(k-q,i\omega_m-x)+f(-x)\Pi_{aa'}(q,i\omega_m-x)A_G^{aa'}(k-q,x)\Big].
\eean
Analytically continued onto the real frequency axis, the self-energy becomes
\ben
\Sigma^{(2)}_{aa'}(k,\omega+i\eta)=\frac{J_K^2}{2N}\sum_q \int \frac{dx}{2\pi} \Big[ n_B(x)A_{\Pi}^{aa'}(q,x)G_{aa'}(k-q,\omega+i\eta-x)+f(-x)\Pi_{aa'}(q,\omega-x+i\eta)A_G^{aa'}(k-q,x)\Big].
\een

\subsection{Saddle point equations for the Haldane model}\label{sec:haldane}
In the Haldane model, the bi-partite lattice is needed so here the self-energy saddle point equations are derived for bi-partite lattice. The Green's function matrix in the Lindhard electron-hole pairs is
\ben
g(k,z)=\frac{\bb P^+(k)}{z-\eps_k}+\frac{\bb P^-(k)}{z+\eps_k}, \hspace{2mm} P^{\pm}(k)=\frac{1\pm \bb N(k)}{2}
\een
where
\ben
N(k)=\frac{1}{\eps_k}\left[\begin{array}{cc}
2 t_2 \sum_i \sin(k_i\cdot b_i)&-t_1(1+e^{-ik\cdot a_1}+e^{-ik\cdot a_2})\\
-t_1(1+e^{ik\cdot a_1}+e^{ik\cdot a_2})&-2 t_2 \sum_i \sin(k_i\cdot b_i)\end{array}\right].
\een
Here $a_1=(3/2,\sqrt{3}/2)$, $a_2=(3/2,-\sqrt{3}/2)$ are the unit vectors and $b_1=(0,-\sqrt{3})$, $b_2=(3/2,\sqrt{3}/2)$ and $b_3=(-3/2,\sqrt{3}/2)$ are the second nearest neighbor vectors of Haldane term.
And
$
\eps_k^2=[2 t_2 \sum_i \sin(k_i\cdot b_i)]^2+\abs{t_1(1+e^{-ik\cdot a_1}+e^{-ik\cdot a_2})}^2.
$
The (low frequency) saddle point equation as derived in the past section is
\be
\sigma_{aa'}(k,i\eta)=\frac{J_K^2}{2N}\sum_q \int dx [n(-x)A^{\Pi}_{aa'}(q,-x)G_{aa'}(k-q,x+i\eta)+f(-x)\Pi_{aa'}(q,-x+i\eta)A^G_{aa'}(k-q,x)]\label{sigH}
\ee
where 
\bean
\Pi_{aa'}(q,z)=&&\frac{1}{N}\sum_p \Big [ P^+_{aa'}(p+q)P^+_{a'a}(p)\frac{f(\eps_p)-f(\eps_{p+q})}{z+\eps_p-\eps_{p+q}}+P^+_{aa'}(p+q)P^-_{a'a}(p)\frac{f(-\eps_p)-f(\eps_{p+q})}{z-\eps_p-\eps_{p+q}}\\
&&+P^-_{aa'}(p+q)P^+_{a'a}(p)\frac{f(\eps_p)-f(-\eps_{p+q})}{z+\eps_p+\eps_{p+q}}+P^+_{aa'}(p+q)P^+_{a'a}(p)\frac{f(-\eps_p)-f(-\eps_{p+q})}{z-\eps_p+\eps_{p+q}}
\Big].
\eean
In the zero temperature, the first term and the last term vanish because of the Fermi function, giving the result of $\Pi_{aa'}$ as
\ben
\Pi_{aa'}(q,z)=\frac{1}{N}\sum_p \Big [P^+_{aa'}(p+q)P^-_{a'a}(p)\frac{1}{z-\eps_p-\eps_{p+q}}-P^-_{aa'}(p+q)P^+_{a'a}(p)\frac{1}{z+\eps_p+\eps_{p+q}} \Big]
\een
and
Green's function
\ben
G(k,z)=\frac{1}{2}\Big[\frac{1+\bb M(k)}{z-E_k}+\frac{1-\bb M(k)}{z+E_k} \Big]
, \qquad 
\bb M(k)=\frac{1}{E_k}\left[\begin{array}{cc}
\sigma_3&\sigma_-\\
\sigma_+&-\sigma_3
\end{array}\right], \qquad 
E_k=\sqrt{\sigma^2_3+\abs{\sigma_+}^2}.
\een
Here we assume the form of the $\bb M(k)$ matrix because of the Hamiltonian of Haldane Honeycomb model. The contributed poles of the above $\Pi_{aa'}$ and $G_{aa'}$ come from $\delta(\omega+\eps_p+\eps_{p+q})$ and $\delta(\omega-E_k)$ such that the spectral functions are
\ben
A^{\Pi}_{aa'}(q,\omega)=-\frac{2\pi}{4N}\sum_p [\delta_{aa'}-N_{aa'}(p+q)][\delta_{aa'}+N_{aa'}(p)]\delta(\omega+\eps_p+\eps_{p+q})
\een
and $A^G_{aa'}(k)=\pi [1+M_{aa'}(k)]\delta(\omega-E_k)$. After plugging these spectral functions in the Eq.\,\pref{sigH}, the real part of the saddle point equation becomes
\bean
\Re\sigma_{aa'}(k)=\frac{J_K^2}{16}\sum_{pq}&&\Big\{f(-E_{k-q})[\delta_{aa'}+M_{aa'}(k-q)]\Big[\frac{[\delta_{aa'}-N_{aa'}(p+q)][\delta_{a'a}+N_{a'a}(p)]}{E_{k-q}-\eps_p-\eps_{p+q}}-\frac{[\delta_{aa'}+N_{aa'}(p+q)][\delta_{a'a}-N_{a'a}(p)]}{E_{k-q}+\eps_p+\eps_{p+q}}\Big]\\
&&-n(-\eps_p-\eps_{p+q})\Big[-\frac{\delta_{aa'}+M_{aa'}(k-q)}{E_{k-q}-\eps_p-\eps_{p+q}}+\frac{\delta_{aa'}-M_{aa'}(k-q)}{E_{k-q}+\eps_p+\eps_{p+q}} \Big][\delta_{aa'}-N_{aa'}(p+q)][\delta_{a'a}+N_{a'a}(p)]\Big\}.
\eean
In the zero temperature limit, the equation will be simplified further
\bean
&&\Re\sigma_{aa'}(k)=\\
&&-\frac{J_K^2}{16N^2}\sum_{pq}\frac{[\delta_{aa'}+M_{aa'}(k-q)][\delta_{aa'}+N_{aa'}(p+q)][\delta_{a'a}-N_{a'a}(p)]-[\delta_{aa'}-M_{aa'}(k-q)][\delta_{aa'}-N_{aa'}(p+q)][\delta_{a'a}+N_{a'a}(p)]}{E_{k-q}+\eps_p+\eps_{p+q}}\\
&&+\frac{J_K^2}{16N^2}\sum_{pq}\frac{[\delta_{aa'}+M_{aa'}(k-q)][\delta_{aa'}-N_{aa'}(p+q)][\delta_{a'a}+N_{a'a}(p)]-[\delta_{aa'}+M_{aa'}(k-q)][\delta_{aa'}-N_{aa'}(p+q)][\delta_{a'a}+N_{a'a}(p)]}{E_{k-q}-\eps_p-\eps_{p+q}}\\
&&=-\frac{J_K^2}{8N^2}\sum_{pq}\Big\{\frac{\delta_{aa'}[N_{aa'}(p+q)-N_{a'a}(p)]}{E_{k-q}+\eps_p+\eps_{p+q}}+\frac{M_{aa'}(k-q)[\delta_{aa'}-N_{aa'}(p+q)N_{a'a}(p)]}{E_{k-q}+\eps_p+\eps_{p+q}}\Big\}.
\eean

\subsection{Lifetime of Spinons and Phase Transition}\label{sec:lifetime}
The saddle point equation which have been used to study the one-dimensional system is following
\bean
\sigma_k(i\eta)&=&\frac{J_K^2}{N^2}\sum_{q}\int \frac{dx}{2\pi} \Big[n(x)A^{\Pi}_{q}(x)G_{k-q}(i\eta-x)+f(-x)\Pi_q(i\eta-x)A_{k-q}^G(x) \Big]\\
&=&\frac{J_K^2}{2N^2}\sum_{k',q}  \Big[ n_B(\eps_{q}-\eps_{k'})\frac{f_{k'}-f_{q}}{\eps_{k'}-\eps_{q}-\sigma_{k-q}+i\eta}+\underbrace{\int \frac{dx}{\pi} f(-x)\frac{f_{k'}-f_{q}}{-x+\eps_{k'}-\eps_{q}+i\eta}\frac{-\sigma_{k-q}''}{(x-\sigma_{k-q}')^2+\sigma_{k-q}''^2}}_{\sigma^{(2)}_{k,q}}\Big]\\
\eean
where the second part with frequency integral is defined as $\sigma^{(2)}$. One can make this integration as a complex integration
\bean
\sigma^{(2)}_{k,q}&=&\oint \frac{dz}{\pi} f(-z)\frac{f_{k'}-f_{q}}{z-(\eps_{k'}-\eps_{q}+i\eta)}\frac{\sigma_{k-q}''}{(z-\sigma_{k-q})(z-\sigma_{k-q}^*)}\\
&=&I_1+I_2+I_3.
\eean
There are three different poles $z_1=\eps_{k'}-\eps_{q}+i\eta$, $z_2=\sigma_{k-q}^*$ and $z_3=i\omega_n$, 
where the third poles contain all the Fermionic Matsubara frequencies. Here $I_1$ and $I_2$ are
\ben
I_1=f(\eps_{q}-\eps_{k'})(f_{k'}-f_{q})\Big[\frac{1}{\eps_{k'}-\eps_{q}+i\eta-\sigma_{k-q}}-\frac{1}{\eps_{k'}-\eps_{q}+i\eta-\sigma_{k-q}^*}\Big], \qquad
I_2=f(-\sigma_{k-q}^*)\frac{f_{k'}-f_{q}}{\eps_{k'}-\eps_{q}+i\eta-\sigma_{k-q}^*}
\een
and $I_3$ is
\bean
I_3&=&\frac{1}{\beta}\sum_{n\geq 0}\frac{f_{k'}-f_{q}}{i\omega_n-(\eps_{k'}-\eps_{q}+i\eta)}\frac{2i\sigma_{k-q}''}{(i\omega_n-\bar{\sigma})(i\omega_n-\sigma_{k-q}^*)}\\
&=&\frac{1}{2\pi i}\frac{f_{k'}-f_{q}}{\eps_{k'}-\eps_{q}-\bar{\sigma}}\Big[\Psi(\frac{1}{2}+i\frac{\sigma_{k-q}}{2\pi T})-\Psi(\frac{1}{2}+i\frac{\eps_{k'}-\eps_{q}}{2\pi T}) \Big]-\frac{1}{2\pi i}\frac{f_{k'}-f_{q}}{\eps_{k'}-\eps_{q}-\sigma_{k-q}^*}\Big[\Psi(\frac{1}{2}+i\frac{\sigma_{k-q}^*}{2\pi T})-\Psi(\frac{1}{2}+i\frac{\eps_{k'}-\eps_{q}}{2\pi T}) \Big]\\
\eean
where the property of digamma function $\Psi(\frac{1}{2}+a)-\Psi(\frac{1}{2}+b)=\sum_{n=0}^{\infty}\left(\frac{1}{n+b}-\frac{1}{n+a}\right)$ has been used. Then combine everything we will get the final form we need 
\bean
\sigma_{k}&=&\sum_{k,q}\frac{(f_k-f_q)\Big[n(\eps_q-\eps_k)+f(\eps_q-\eps_k) \Big]}{(\eps_k-\eps_q)-\sigma_{k-q}}+\frac{1}{\pi}\sum_{k,q}\text{Im}\Big[\frac{f_k-f_q}{\eps_k-\eps_q-\sigma_{k-q}}\Big(\Psi(\frac{1}{2}+\frac{i\sigma_{k-q}}{2\pi T})-\Psi(\frac{1}{2}+i\frac{\eps_k-\eps_q}{2\pi T})\Big) \Big].\\
\eean

In the localized phase, $\sigma_{k-q}=\bar{\sigma}$ which doesn't have any momentum dependence. Then the second part vanishs, in half filling case, where $\sigma'+\lambda=0$ giving $\sigma=i\sigma''$, as following:
\bean
\sigma^{(2)}&=&\frac{1}{2\pi i}\sum_{k,q}\Big[\frac{f_k-f_q}{\eps_k-\eps_q-i\sigma''}\Big(\Psi(\frac{1}{2}-\frac{\sigma''}{2\pi T})-\Psi(\frac{1}{2}+i\frac{\eps_k-\eps_q}{2\pi T})\Big) -\frac{f_k-f_q}{\eps_k-\eps_q+i\sigma''}\Big(\Psi(\frac{1}{2}-\frac{\sigma''}{2\pi T})-\Psi(\frac{1}{2}-i\frac{\eps_k-\eps_q}{2\pi T})\Big) \Big]\\
&=&\frac{1}{2\pi i}\sum_{k,q}\Big[\frac{f_k-f_q}{\eps_k-\eps_q-i\sigma''}\Big(\Psi(\frac{1}{2}-\frac{\sigma''}{2\pi T})-\Psi(\frac{1}{2}+i\frac{\eps_k-\eps_q}{2\pi T})\Big) -\frac{f_q-f_k}{\eps_q-\eps_k+i\sigma''}\Big(\Psi(\frac{1}{2}-\frac{\sigma''}{2\pi T})-\Psi(\frac{1}{2}+i\frac{\eps_k-\eps_q}{2\pi T})\Big) \Big]\\
&=&0.
\eean
In second part of the second line, $k$ and $q$ are switched. The self-energy becomes
\bean
\bar{\sigma}&=&\sum_{k,q}\frac{(f_k-f_q)\Big[n(\eps_q-\eps_k)+f(\eps_q-\eps_k) \Big]}{(\eps_k-\eps_q)-\bar{\sigma}}.
\eean
The real part of $\sigma$ is zero and only the imaginary part survives which makes this equation the same as Eq.\,\pref{eqlife}.
\ew

\subsection{Delocalization Transitons in Static RKKY case}
The saddle point equation used throughout the paper is
\ben
\Sigma_k(\tau)=\frac{J_K^2}{2N}\sum_q G_{k-q}(\tau)\Pi_q(\tau).
\een
In the static limit from Eq.\,\pref{static1}, this simplifies to:
\ben
\Sigma_k(0^-)=\frac{J_K^2}{2N}\sum_q G_{k-q}(0^-)\Pi_q,
\een
where $\Pi_q$ is the static (zero frequency) susceptibility of electrons 
\bean
\Pi_q&=&-\frac{1}{N}\sum_p\frac{f_p-f_{p+q}}{\eps_p-\eps_{p+q}}\\
\eean
and the equal time Green's function is defined as:
\bean
G_{k-q}(0^-)&=&\frac{1}{\beta}\sum_n e^{-i \omega_n 0^-} G_{k-q}(i\omega_n)\\
&=&-\oint \frac{dz}{2\pi i} f(z) \frac{1}{z-\sigma_{k-q}}\\
&=&f(\sigma_{k-q})\\
&\approx&\frac{1}{2}-\frac{1}{4T}\sigma_{k-q}.
\eean
In the last line we have expanded the Fermi function with $\sigma_{k-q}$ assuming that it is small. The saddle point equation describing the system can be expressed as
\ben
\bar{\sigma}+\sigma_{k}=\frac{J_K^2}{2N}\sum_q (\frac{1}{2}-\frac{1}{4T}\sigma_{k-q})\Pi_q.
\een
Here, the first term on each side of the equation corresponds to the localized spin configuration, where $\bar{\sigma}=\frac{J_K^2}{4N}\sum_q \Pi_q$. This describes the system when the spins are localized, representing a static, non-dispersive state. The second term, which describes the delocalized scenario, reflects the dispersion of spinons and can be written as:
\ben
\sigma_k=\frac{J_K^2}{8TN}\sum_q \Pi_{k-q}\sigma_{q}
\een
and transform to the (arbitrary dimensional) real space 
\bean
\sigma(\vec r)&=&\frac{J_K^2}{8TN^2}\sum_{k,q}e^{i\vec k\cdot\vec r} \Pi_{\vec k-\vec q}\sigma_{\vec q}\\
&=&\frac{J_K^2}{8TN^4}\sum_{k,q}\sum_{\vec r_1,\vec r_2} \Pi(\vec r_1)\sigma(\vec r_2) e^{i\vec k\cdot \vec r-i(\vec k-\vec q)\cdot \vec r_1-i\vec q\cdot\vec r_2}\\
&=&\frac{J_K^2}{8T} \Pi(\vec r)\sigma(\vec r).
\eean
Here, $\vec r$ represents the real-space coordinate on the lattice, while $\Pi(\vec r)$ denotes the static RKKY interaction in real space, which exhibits oscillatory behavior as $\abs{\vec r}$ varies. This expression indicates how spinons delocalize and interact across the lattice, with the temperature $T$ playing a critical role in the dynamics. The polarization function $\Pi_q$ serves as a key factor in determining how interactions are mediated, particularly through the RKKY mechanism. In essence, the interplay between localized and delocalized behaviors is governed by the Kondo coupling $J_K$, the temperature, and the system's susceptibility encapsulated in $\Pi_q$.

\bibliography{KL}

\end{document}